\documentclass[conference]{IEEEtran}

\usepackage{tikz}
\usepackage{pgfplots}
\usepackage{bm}
\usepackage{amsmath}
\pgfplotsset{compat=1.16}

\DeclareMathOperator{\sgn}{sgn}
\DeclareMathOperator*{\argmin}{argmin}
\DeclareMathOperator*{\atanh}{atanh}

\begin{document}

\bstctlcite{IEEEexample:BSTcontrol}

\title{Successive Syndrome-Check Decoding of\\Polar Codes}

\author{
\IEEEauthorblockN{Seyyed Ali Hashemi\IEEEauthorrefmark{1}, Marco Mondelli\IEEEauthorrefmark{2}, John Cioffi\IEEEauthorrefmark{3}, Andrea Goldsmith\IEEEauthorrefmark{4}}

\IEEEauthorblockA{\IEEEauthorrefmark{1}Qualcomm, USA}
\IEEEauthorblockA{\IEEEauthorrefmark{2}Institute of Science and Technology, Austria}
\IEEEauthorblockA{\IEEEauthorrefmark{3}Department of Electrical Engineering, Stanford University, USA}
\IEEEauthorblockA{\IEEEauthorrefmark{4}Department of Electrical and Computer Engineering, Princeton University, USA}
\IEEEauthorblockA{hashemi@qti.qualcomm.com,
marco.mondelli@ist.ac.at,
cioffi@stanford.edu,
goldsmith@princeton.edu}
}
\maketitle

\begin{abstract}
A two-part successive syndrome-check decoding of polar codes is proposed with the first part successively refining the received codeword and the second part checking its syndrome. A new formulation of the successive-cancellation (SC) decoding algorithm is presented that allows for successively refining the received codeword by comparing the log-likelihood ratio value of a frozen bit with its predefined value. The syndrome of the refined received codeword is then checked for possible errors. In case there are no errors, the decoding process is terminated. Otherwise, the decoder continues to refine the received codeword. The proposed method is extended to the case of SC list (SCL) decoding by terminating the decoding process when the syndrome of the best candidate in the list indicates no errors. Simulation results show that the proposed method reduces the time-complexity of SC and SCL decoders and their fast variants, especially at high signal-to-noise ratios.
\end{abstract}

\IEEEpeerreviewmaketitle

\section{Introduction}

Polar codes are a class of coding schemes that can provably achieve the capacity of a binary-input memoryless symmetric channel with low-complexity encoding and decoding algorithms as the code length tends to infinity \cite{arikan2009}.
The successive-cancellation (SC) decoder for a polar code of length $N$, with which polar codes can achieve capacity, has complexity $O(N\log N)$. The SC list (SCL) decoding with list size $L>1$ improves the error rate performance of polar codes at finite block lengths and has complexity $O(LN\log N)$ \cite{tal2015list}. SC and SCL decoding algorithms are serial in nature in the sense that the decoders progress by decoding the bits one by one. Consequently, the latency of SC and SCL decoders is high, which adversely affects their practical implementation.

Several attempts have been made to improve the speed of SC and SCL decoders for polar codes. In particular, the decoding process is represented as message passing on a binary tree where the messages are either soft log-likelihood ratio (LLR) values or hard bit estimations. Since the calculation of soft messages is more costly than hard bit estimations, the decoding latency is modeled as the number of edges in the binary tree that need to be traversed by the underlying decoder to calculate the soft messages. The goal of fast SC and SCL decoders is to prune the decoding tree to reduce the number of edges. This pruning is achieved by the following main methods:
\begin{enumerate}
    \item Identifying specific constituent codes of polar codes that can be decoded efficiently without traversing the decoding tree \cite{BYuan2Bit,sarkis2013increasing,husmann2017reduced,alamdar2011simplified,sarkis2014fast,giard2016multi,hanif2017fast,condo2018generalized,gamage2019low,ercan2017reduced,ercan2019operation,sarkis2016fast,hashemi2016fast,hashemi2017fast,hanif2018fast,mondelli2021sublinear,hashemi2021parallelism};
    \item Checking the syndrome of the constituent codes of polar codes to avoid traversing the decoding tree when the syndrome check is satisfied \cite{yoo2016comml,choi2017comml,kim2018ET};
    \item Adjusting the code to increase the number of specific constituent codes of polar codes \cite{huang2012latency,balatsoukas2014enabling,zhang2015simplified,638mbps,giard2018fast};
    \item Comparing the error probability of constituent codes of polar codes with a threshold to avoid traversing the decoding tree when the error probability is small \cite{li2018low,zheng2020icc}.
\end{enumerate}
The first two methods are lossless in the sense that they do not result in error-correction performance degradation with respect to the conventional SC and SCL decoders, while the other two methods may cause error-correction performance loss.

In this paper, SC decoding is reformulated from a novel perspective that allows for successively refining the codeword that was received from the channel. It is shown that this codeword refinement strategy can only take place when a frozen bit that has a predefined value is decoded. Thus, there is no need to calculate soft messages for information bits. Moreover, the refinement of the received codeword is only necessary for those frozen bits whose LLR values do not match their predefined values. Using the equivalence of soft and hard decoding when there is no error, a successive syndrome-check decoder is presented that significantly speeds up the decoding process, especially when the transmission channel experiences low levels of noise.

\section{Preliminaries}

\subsection{Polar Codes} \label{sec:PC}

Let $K$ denote the number of information bits and $N=2^n$ denote the block length of the code. A polar code is represented by $\mathcal{P}\left(N,K\right)$ that has rate $R = \frac{K}{N}$. The encoded vector $\bm{x} = \{x_0,x_1,\ldots,x_{N-1}\}$ is generated by the multiplication of the input vector $\bm{u} = \{u_0,u_1,\ldots,u_{N-1}\}$ and the generator matrix $\bm{G}^{\otimes n}$, and it is expressed as $\bm{x}=\bm{u}\bm{G}^{\otimes n}$. The generator matrix is constructed by the $n$-th Kronecker power of the matrix $\bm{G}=\left[\begin{smallmatrix}1&0\\1&1\end{smallmatrix}\right]$.

The input vector $\bm{u}$ is divided into two sets: the set of $K$ information bits $\mathcal{I}$, and the set of $N-K$ frozen bits $\mathcal{F}$ whose values are known to the receiver. Without loss of generality, in this paper, all frozen bits are set to zero \cite{arikan2009}. Furthermore, the codeword $\bm{x}$ is modulated with binary phase-shift keying (BPSK) modulation that maps $\left\{0,1\right\}$ to $\left\{+1,-1\right\}$. The modulated signal is then transmitted through an additive white Gaussian noise (AWGN) channel. The received signal is transformed into log-likelihood ratio (LLR) values $\bm{y} = \{y_0,y_1,\ldots,y_{N-1}\}$ that is then processed for decoding. 


\subsection{Successive-Cancellation Decoding}

\begin{figure}
    \centering
    \begin{tikzpicture}[scale=.6, thick]

  \draw[very thin,gray,dashed] (1,.5) -- (1,-7.5);
  \draw[very thin,gray,dashed] (3,.5) -- (3,-7.5);
  \draw[very thin,gray,dashed] (5,.5) -- (5,-7.5);
  \draw[very thin,gray,dashed] (7,.5) -- (7,-7.5);


  \node at (.5,0) {$0$};
  \node at (.5,-1) {$0$};
  \node at (.5,-2) {$0$};
  \node at (.5,-3) {$u_3$};
  \node at (.5,-4) {$0$};
  \node at (.5,-5) {$u_5$};
  \node at (.5,-6) {$u_6$};
  \node at (.5,-7) {$u_7$};

  \draw (1,0) -- ++(.75,0);
  \draw (2.25,0) -- ++(.75,0);
  \draw (1,-1) -- ++(.75,0);
  \draw (2.25,-1) -- ++(.75,0);

  \draw (2,0) circle [radius=.25] node {\tiny $f$};
  \draw (2,-1) circle [radius=.25] node {\tiny $g$};
  
  \draw (3,-1) -- ++(-.75,1);
  \draw (3,0) -- ++(-.75,-1);
  
  \draw (1.75,-1.5) -- ++(.25,0) -- ++(0,.25);

  \draw (1,-2) -- ++(.75,0);
  \draw (2.25,-2) -- ++(.75,0);
  \draw (1,-3) -- ++(.75,0);
  \draw (2.25,-3) -- ++(.75,0);

  \draw (2,-2) circle [radius=.25] node {\tiny $f$};
  \draw (2,-3) circle [radius=.25] node {\tiny $g$};
  
  \draw (3,-3) -- ++(-.75,1);
  \draw (3,-2) -- ++(-.75,-1);
  
  \draw (1.75,-3.5) -- ++(.25,0) -- ++(0,.25);

  \draw (1,-4) -- ++(.75,0);
  \draw (2.25,-4) -- ++(.75,0);
  \draw (1,-5) -- ++(.75,0);
  \draw (2.25,-5) -- ++(.75,0);

  \draw (2,-4) circle [radius=.25] node {\tiny $f$};
  \draw (2,-5) circle [radius=.25] node {\tiny $g$};
  
  \draw (3,-5) -- ++(-.75,1);
  \draw (3,-4) -- ++(-.75,-1);
  
  \draw (1.75,-5.5) -- ++(.25,0) -- ++(0,.25);

  \draw (1,-6) -- ++(.75,0);
  \draw (2.25,-6) -- ++(.75,0);
  \draw (1,-7) -- ++(.75,0);
  \draw (2.25,-7) -- ++(.75,0);

  \draw (2,-6) circle [radius=.25] node {\tiny $f$};
  \draw (2,-7) circle [radius=.25] node {\tiny $g$};
  
  \draw (3,-7) -- ++(-.75,1);
  \draw (3,-6) -- ++(-.75,-1);
  
  \draw (1.75,-7.5) -- ++(.25,0) -- ++(0,.25);


  \draw (3,0) -- ++(.75,0);
  \draw (4.25,0) -- ++(.75,0);
  \draw (3,-2) -- ++(.75,0);
  \draw (4.25,-2) -- ++(.75,0);

  \draw (4,0) circle [radius=.25] node {\tiny $f$};
  \draw (4,-2) circle [radius=.25] node {\tiny $g$};
  
  \draw (5,-2) -- ++(-.75,2);
  \draw (5,0) -- ++(-.75,-2);
  
  \draw (3.75,-2.5) -- ++(.25,0) -- ++(0,.25);

  \draw (3,-1) -- ++(.75,0);
  \draw (4.25,-1) -- ++(.75,0);
  \draw (3,-3) -- ++(.75,0);
  \draw (4.25,-3) -- ++(.75,0);

  \draw (4,-1) circle [radius=.25] node {\tiny $f$};
  \draw (4,-3) circle [radius=.25] node {\tiny $g$};
  
  \draw (5,-3) -- ++(-.75,2);
  \draw (5,-1) -- ++(-.75,-2);
  
  \draw (3.75,-3.5) -- ++(.25,0) -- ++(0,.25);

  \draw (3,-4) -- ++(.75,0);
  \draw (4.25,-4) -- ++(.75,0);
  \draw (3,-6) -- ++(.75,0);
  \draw (4.25,-6) -- ++(.75,0);

  \draw (4,-4) circle [radius=.25] node {\tiny $f$};
  \draw (4,-6) circle [radius=.25] node {\tiny $g$};
  
  \draw (5,-6) -- ++(-.75,2);
  \draw (5,-4) -- ++(-.75,-2);
  
  \draw (3.75,-6.5) -- ++(.25,0) -- ++(0,.25);

  \draw (3,-5) -- ++(.75,0);
  \draw (4.25,-5) -- ++(.75,0);
  \draw (3,-7) -- ++(.75,0);
  \draw (4.25,-7) -- ++(.75,0);

  \draw (4,-5) circle [radius=.25] node {\tiny $f$};
  \draw (4,-7) circle [radius=.25] node {\tiny $g$};
  
  \draw (5,-7) -- ++(-.75,2);
  \draw (5,-5) -- ++(-.75,-2);
  
  \draw (3.75,-7.5) -- ++(.25,0) -- ++(0,.25);


  \draw (5,0) -- ++(.75,0);
  \draw (6.25,0) -- ++(.75,0);
  \draw (5,-4) -- ++(.75,0);
  \draw (6.25,-4) -- ++(.75,0);

  \draw (6,0) circle [radius=.25] node {\tiny $f$};
  \draw (6,-4) circle [radius=.25] node {\tiny $g$};
  
  \draw (7,-4) -- ++(-.75,4);
  \draw (7,0) -- ++(-.75,-4);
  
  \draw (5.75,-4.5) -- ++(.25,0) -- ++(0,.25);

  \draw (5,-1) -- ++(.75,0);
  \draw (6.25,-1) -- ++(.75,0);
  \draw (5,-5) -- ++(.75,0);
  \draw (6.25,-5) -- ++(.75,0);

  \draw (6,-1) circle [radius=.25] node {\tiny $f$};
  \draw (6,-5) circle [radius=.25] node {\tiny $g$};
  
  \draw (7,-5) -- ++(-.75,4);
  \draw (7,-1) -- ++(-.75,-4);
  
  \draw (5.75,-5.5) -- ++(.25,0) -- ++(0,.25);

  \draw (5,-2) -- ++(.75,0);
  \draw (6.25,-2) -- ++(.75,0);
  \draw (5,-6) -- ++(.75,0);
  \draw (6.25,-6) -- ++(.75,0);

  \draw (6,-2) circle [radius=.25] node {\tiny $f$};
  \draw (6,-6) circle [radius=.25] node {\tiny $g$};
  
  \draw (7,-6) -- ++(-.75,4);
  \draw (7,-2) -- ++(-.75,-4);
  
  \draw (5.75,-6.5) -- ++(.25,0) -- ++(0,.25);

  \draw (5,-3) -- ++(.75,0);
  \draw (6.25,-3) -- ++(.75,0);
  \draw (5,-7) -- ++(.75,0);
  \draw (6.25,-7) -- ++(.75,0);

  \draw (6,-3) circle [radius=.25] node {\tiny $f$};
  \draw (6,-7) circle [radius=.25] node {\tiny $g$};
  
  \draw (7,-7) -- ++(-.75,4);
  \draw (7,-3) -- ++(-.75,-4);
  
  \draw (5.75,-7.5) -- ++(.25,0) -- ++(0,.25);

  \node at (7.5,0) {$y_0$};
  \node at (7.5,-1) {$y_1$};
  \node at (7.5,-2) {$y_2$};
  \node at (7.5,-3) {$y_3$};
  \node at (7.5,-4) {$y_4$};
  \node at (7.5,-5) {$y_5$};
  \node at (7.5,-6) {$y_6$};
  \node at (7.5,-7) {$y_7$};

\end{tikzpicture}
    \begin{tikzpicture}[scale=.93, thick]

  \draw[very thin,gray,dashed] (1,.5) -- (1,-1.5);
  \draw[very thin,gray,dashed] (3,.5) -- (3,-1.5);


  \node at (.5,0) {$u_0,l_0$};
  \node at (.5,-1) {$u_1,l_1$};

  \draw (1,0) -- (1.75,0);
  \draw (2.25,0) -- (3,0);
  \draw (1,-1) -- (1.75,-1);
  \draw (2.25,-1) -- (3,-1);

  \draw (2,0) circle [radius=.25] node {$f$};
  \draw (2,-1) circle [radius=.25] node {$g$};
  
  \draw (3,-1) -- (2.25,0);
  \draw (3,0) -- (2.25,-1);
  
  \draw (1.75,-1.5) -- ++(.25,0) -- ++(0,.25);
  
  \node at (1.5,-1.5) {$\hat{u}_0$};

  \node at (3.5,0) {$x_0,y_0$};
  \node at (3.5,-1) {$x_1,y_1$};

\end{tikzpicture}
    \caption{SC decoding on the factor graph of $\mathcal{P}(8,4)$ with $\mathcal{I} = \{u_3,u_5,u_6,u_7\}$ and $\mathcal{F} = \{u_0,u_1,u_2,u_4\}$ (left), and the building block of the SC decoder (right).}
    \label{fig:decGraph}
\end{figure}
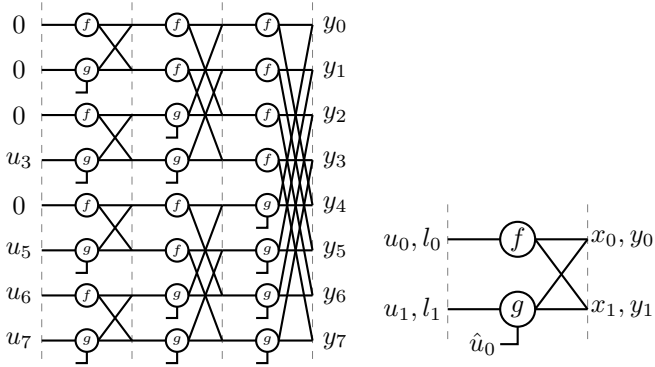

SC decoding can be represented on a factor graph as illustrated in Fig.~\ref{fig:decGraph} for $\mathcal{P}(8,4)$ with $\mathcal{I} = \{u_3,u_5,u_6,u_7\}$ and $\mathcal{F} = \{u_0,u_1,u_2,u_4\}$. The factor graph has $n+1$ levels and each decoding step consists of traversing the factor graph by performing $f$ and $g$ functions on the building block of the SC decoder (see Fig.~\ref{fig:decGraph}) as
\begin{align}
    l_0 = f(y_0,y_1) &= 2 \atanh\left(\tanh\left(\frac{y_0}{2}\right)\tanh\left(\frac{y_1}{2}\right)\right) \\
    &\approx \sgn(y_0y_1)\min(|y_0|,|y_1|) \text{,} \label{eq:fFunction} \\
    \hat{u}_0 &= 0 \text{ if } l_0>0 \text{ or } u_0\in\mathcal{F} \text{ else } 1 \text{,} \label{eq:fu0} \\
    l_1 = g(y_0,y_1,\hat{u}_0) &= y_1 + (-1)^{\hat{u}_0}y_0 \text{,} \label{eq:gFunction} \\
    \hat{u}_1 &= 0 \text{ if } l_1>0 \text{ or } u_1\in\mathcal{F} \text{ else } 1 \text{.} \label{eq:gu1}
\end{align}
Note that after performing the $f$ function, the bit $u_0$ is estimated as $\hat{u}_0$, which is an input to the $g$ function. Finally, when $u_1$ is estimated as $\hat{u}_1$, the \emph{partial sums}, $x_0$ and $x_1$, can be calculated as
\begin{align}
    x_0 &= \hat{u}_0 \oplus \hat{u}_1 \text{,} \label{eq:x0} \\
    x_1 &= \hat{u}_1 \text{.} \label{eq:x1}
\end{align}

SC decoding has strong data dependency in the sense that the decoding of each bit $u_i$ requires the estimation of all the previous bits $\{u_0,u_1,\ldots,u_{i-1}\}$. In a fully-parallel implementation of the SC decoder \cite{arikan2009}, all the $f$ and $g$ functions that can be performed in parallel are grouped together. Thus, the latency is modeled as the number of times the groups of functions in (\ref{eq:fFunction}) and (\ref{eq:gFunction}) are performed to move from one level of the factor graph to another. This is due to the fact that the bit-wise calculations in (\ref{eq:x0}) and (\ref{eq:x1}) are of lower complexity than calculating the LLR values.

SCL decoding \cite{tal2015list} improves the error-correction performance of SC decoding by keeping a list of the most likely candidates at each decoding step. Instead of estimating each information bit as either $0$ or $1$ as in (\ref{eq:fu0}) and (\ref{eq:gu1}), SCL decoding considers both possibilities $0$ and $1$. To avoid an exponential growth in the number of candidates, SCL decoding keeps the $L$ most likely ones based on calculating a path metric. The latency of SCL decoding can be modelled similar to that of SC decoding, with the addition of $K$ time steps to update and sort the path metrics.

\subsection{Syndrome-Check Decoding}

The LLR values that are calculated based on the received message can be readily converted into estimated bits. A parity-check matrix can be used to check if these estimated bits represent a potential transmitted codeword. The idea is to generate a syndrome by multiplying the estimated bits $\bm{x}$ and the parity-check matrix $\bm{H}$, and check if all the parity constraints are satisfied. In case of a mismatch, the estimated bits $\bm{x}$ are refined as $\bm{x} = \bm{x}\oplus\bm{e}$ using an \emph{error vector} $\bm{e}$ and checked again against any parity violations. This process can be performed until $\bm{x}$ passes the syndrome check or a predefined maximum number of syndrome-check attempts is reached. Note that the parity-check matrix of polar codes is formed by selecting the columns of $\bm{G}^{\otimes n}$ that correspond to frozen bits. For $\mathcal{P}(8,4)$ with $\mathcal{I} = \{u_3,u_5,u_6,u_7\}$ and $\mathcal{F} = \{u_0,u_1,u_2,u_4\}$, $\bm{G}^{\otimes 3}$ and $\bm{H}$ are
\begin{equation*}
    \bm{G}^{\otimes 3} = \left[
        \begin{smallmatrix}
            1 & 0 & 0 & 0 & 0 & 0 & 0 & 0\\
            1 & 1 & 0 & 0 & 0 & 0 & 0 & 0\\
            1 & 0 & 1 & 0 & 0 & 0 & 0 & 0\\
            1 & 1 & 1 & 1 & 0 & 0 & 0 & 0\\
            1 & 0 & 0 & 0 & 1 & 0 & 0 & 0\\
            1 & 1 & 0 & 0 & 1 & 1 & 0 & 0\\
            1 & 0 & 1 & 0 & 1 & 0 & 1 & 0\\
            1 & 1 & 1 & 1 & 1 & 1 & 1 & 1\\
        \end{smallmatrix}
        \right] \text{, }
    \bm{H} = \left[
        \begin{smallmatrix}
            1 & 0 & 0 & 0 \\
            1 & 1 & 0 & 0 \\
            1 & 0 & 1 & 0 \\
            1 & 1 & 1 & 0 \\
            1 & 0 & 0 & 1 \\
            1 & 1 & 0 & 1 \\
            1 & 0 & 1 & 1 \\
            1 & 1 & 1 & 1 \\
        \end{smallmatrix}
        \right] \text{.}
\end{equation*}
Chase decoding \cite{chase} and guessing random additive noise decoding \cite{grand} are examples that incorporate syndrome-check decoding.

\section{Successive Syndrome-Check Decoding of Polar Codes}

In this section, SC decoding is reformulated to allow for performing successive syndrome checks. This results in an early-termination scheme for SC decoding that reduces the decoding latency, especially when the transmission channel experiences low levels of noise.

\subsection{Successive Syndrome-Check Decoding}

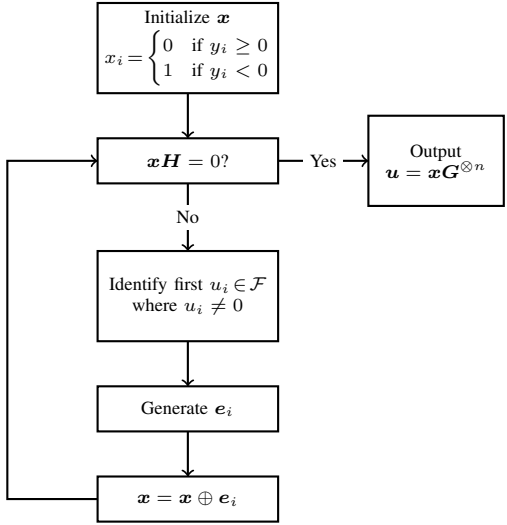
\begin{figure}
    \centering
    \begin{tikzpicture}[scale=1.2,thick]\scriptsize
\draw (0,.5) rectangle ++(2,-1) node [pos=.5,align=center] {Initialize $\bm{x}$\\[.1cm]$x_i \!=\! \begin{cases}0 & \text{if } y_i\geq0 \\1 & \text{if } y_i<0 \end{cases}$};

\draw[->] (1,-.5) -- ++(0,-.5);
\draw (0,-1) rectangle ++(2,-.5) node [pos=.5,align=center] {$\bm{x}\bm{H} = 0$?};

\draw[->] (2,-1.25) -- ++(1,0) node [midway,fill=white] {Yes};
\draw (3,-.75) rectangle ++(1.5,-1) node [pos=.5,align=center] {Output\\$\bm{u} = \bm{x}\bm{G}^{\otimes n}$};

\draw[->] (1,-1.5) -- ++(0,-.75) node [midway,fill=white] {No};
\draw (0,-2.25) rectangle ++(2,-1) node [pos=.5,align=center] {Identify first $u_i\!\in\!\mathcal{F}$\\where $u_i \neq 0$};

\draw[->] (1,-3.25) -- ++(0,-.5);
\draw (0,-3.75) rectangle ++(2,-.5) node [pos=.5,align=center] {Generate $\bm{e}_i$};

\draw[->] (1,-4.25) -- ++(0,-.5);
\draw (0,-4.75) rectangle ++(2,-.5) node [pos=.5,align=center] {$\bm{x} = \bm{x} \oplus \bm{e}_i$};

\draw[->] (0,-5) -- ++(-1,0) -- ++(0,3.75) -- ++(1,0);

\end{tikzpicture}
    \caption{Block diagram of the proposed successive syndrome-check decoder.}
    \label{fig:blockDiagram}
\end{figure}

Fig.~\ref{fig:blockDiagram} shows the block diagram of the proposed successive syndrome-check decoder. The decoder starts by making an initial estimate on the received codeword $\bm{x}$ based on the received LLR values as
\begin{equation}
    x_i = \begin{cases}0 & \text{if } y_i\geq0 \text{,}
    \\1 & \text{if } y_i<0 \text{.} \end{cases} \label{eq:xinit}
\end{equation}
The syndrome of $\bm{x}$ is calculated as $\bm{x}\bm{H}$ and is checked for possible errors. If $\bm{x}\bm{H}=0$, then $\bm{x}$ represents a valid codeword and thus $\bm{u} = \bm{x}\bm{G}^{\otimes n}$ is reported as the decoding result. Otherwise, if $\bm{x}\bm{H}\neq 0$, then there must be an error in the received message. To correct the error, the bit-by-bit decoding schedule of the SC decoder is exploited and the first $u_i \in \mathcal{F}$ is identified whose corresponding value in $\bm{x}\bm{H}$ is not $0$. In fact, starting from $u_0$ and up to the first $u_i \in \mathcal{F}$ where $u_i \neq 0$, any frozen bit whose corresponding value in the syndrome is $0$ does not need to be considered for error correction.

\begin{figure}
    \centering
    \begin{tikzpicture}[scale=.93, thick]

  \draw[very thin,gray,dashed] (1,.5) -- (1,-1.5);
  \draw[very thin,gray,dashed] (3,.5) -- (3,-1.5);


  \node at (.5,0) {$u_0,l_0$};
  \node at (.5,-1) {$u_1,l_1$};

  \draw (1,0) -- (1.75,0);
  \draw (2.25,0) -- (3,0);
  \draw (1,-1) -- (1.75,-1);
  \draw (2.25,-1) -- (3,-1);

  \draw (2,0) circle [radius=.25] node {$f$};
  \draw (2,-1) circle [radius=.25] node {$g$};
  
  \draw (3,-1) -- (2.25,0);
  \draw (3,0) -- (2.25,-1);
  
  \draw (1.75,-1.5) -- ++(.25,0) -- ++(0,.25);
  
  \node at (1.15,-1.5) {$x_0 \!\oplus\! x_1$};

  \node at (3.5,0) {$x_0,y_0$};
  \node at (3.5,-1) {$x_1,y_1$};

\end{tikzpicture}
    \caption{The building block of the proposed successive syndrome-check decoder.}
    \label{fig:SSCdecBB}
\end{figure}
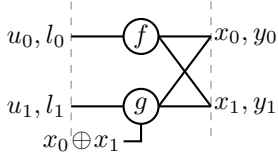

After identifying the first $u_i \in \mathcal{F}$ where $u_i \neq 0$, an error vector $\bm{e}_i$ is generated in an attempt to refine the received message $\bm{x}$. To generate $\bm{e}_i$, the LLR value associated with $u_i$ is calculated by moving from the right to the left of the factor-graph of Fig.~\ref{fig:decGraph}, using a reformulation of SC decoding as depicted in Fig.~\ref{fig:SSCdecBB}. In particular, the $g$ function in the SC decoder (\ref{eq:gFunction}) is replaced with
\begin{equation}
    l_1 = g(y_0,y_1,x_0,x_1) = y_1 + (-1)^{x_0\oplus x_1}y_0 \text{,} \label{eq:gFunctionS}
\end{equation}
where we used the fact that $u_0$ is decoded correctly and $\hat{u}_0 = x_0 \oplus x_1$. In other words, $\bm{x}$ includes all the refinements that result from decoding $u_0$. Note that due to the equivalence of soft and hard decoding when there is no error, the LLR of the first frozen bit whose corresponding estimate in the syndrome is $1$ must be negative, which is in violation of its estimate.

Following the calculation of the LLR value, the error vector $\bm{e}_i$ is generated by moving from the left to the right of the factor-graph of Fig.~\ref{fig:decGraph}. When an $f$ function is encountered, the error follows the path corresponding to the input of $f$ function with the minimum absolute LLR value. When a $g$ function is encountered, the error follows both paths connected to the input of the $g$ function. In Fig.~\ref{fig:SSCdecBB}, the generation of $\bm{e}_i$ translates to
\begin{align}
    f&:
    \begin{cases}
    m = \displaystyle\argmin_{i\in \{0,1\}} |y_i| \text{,} \\
    x_m = 1 \oplus x_m \text{,}
    \end{cases} \label{eq:ferror} \\
    g&:
    \begin{cases}
    x_0 = 1 \oplus x_0 \text{,} \\
    x_1 = 1 \oplus x_1 \text{.}
    \end{cases} \label{eq:gerror}
\end{align}
$\bm{e}_i$ is then used to refine the received codeword $\bm{x}$ as
\begin{equation}
    \bm{x} = \bm{x}\oplus\bm{e}_i \text{.}
\end{equation}
The received codeword is repeatedly updated until its syndrome becomes $0$.

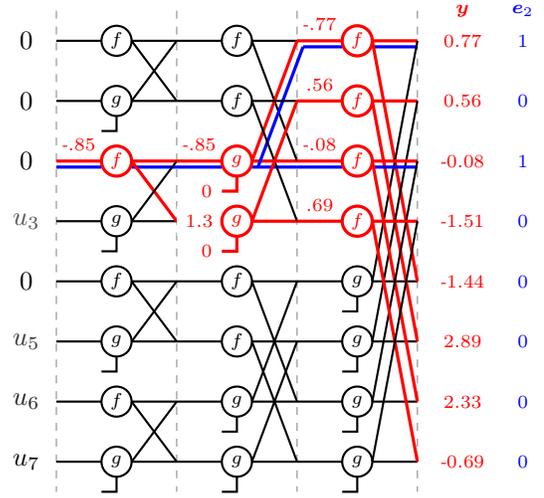
\begin{figure}
    \centering
    \begin{tikzpicture}[scale=.8, thick]

  \draw[very thin,gray,dashed] (1,.5) -- (1,-7.5);
  \draw[very thin,gray,dashed] (3,.5) -- (3,-7.5);
  \draw[very thin,gray,dashed] (5,.5) -- (5,-7.5);
  \draw[very thin,gray,dashed] (7,.5) -- (7,-7.5);


  \node at (.5,0) {\color{black}$0$};
  \node at (.5,-1) {\color{black}$0$};
  \node at (.5,-2) {\color{black}$0$};
  \node at (.5,-3) {\color{black!68.4}$u_3$};
  \node at (.5,-4) {\color{black}$0$};
  \node at (.5,-5) {\color{black!80.9}$u_5$};
  \node at (.5,-6) {\color{black!87.9}$u_6$};
  \node at (.5,-7) {\color{black!99.6}$u_7$};

  \draw (1,0) -- ++(.75,0);
  \draw (2.25,0) -- ++(.75,0);
  \draw (1,-1) -- ++(.75,0);
  \draw (2.25,-1) -- ++(.75,0);

  \draw (2,0) circle [radius=.25] node {\scriptsize $f$};
  \draw (2,-1) circle [radius=.25] node {\scriptsize $g$};
  
  \draw (3,-1) -- ++(-.75,1);
  \draw (3,0) -- ++(-.75,-1);
  
  \draw (1.75,-1.5) -- ++(.25,0) -- ++(0,.25);

  \draw[red,very thick] (1,-2) -- ++(.75,0) node [above,midway] {\scriptsize-$.85$};
  \draw[blue,very thick] (1,-2.1) -- ++(.75,0);
  \draw[red,very thick] (2.25,-2) -- ++(.75,0);
  \draw[blue,very thick] (2.25,-2.1) -- ++(.75,0);
  \draw (1,-3) -- ++(.75,0);
  \draw (2.25,-3) -- ++(.75,0);

  \draw[red,very thick] (2,-2) circle [radius=.25] node {\scriptsize $f$};
  \draw (2,-3) circle [radius=.25] node {\scriptsize $g$};
  
  \draw[red,very thick] (3,-3) -- ++(-.75,1);
  \draw (3,-2) -- ++(-.75,-1);
  
  \draw (1.75,-3.5) -- ++(.25,0) -- ++(0,.25);

  \draw (1,-4) -- ++(.75,0);
  \draw (2.25,-4) -- ++(.75,0);
  \draw (1,-5) -- ++(.75,0);
  \draw (2.25,-5) -- ++(.75,0);

  \draw (2,-4) circle [radius=.25] node {\scriptsize $f$};
  \draw (2,-5) circle [radius=.25] node {\scriptsize $g$};
  
  \draw (3,-5) -- ++(-.75,1);
  \draw (3,-4) -- ++(-.75,-1);
  
  \draw (1.75,-5.5) -- ++(.25,0) -- ++(0,.25);

  \draw (1,-6) -- ++(.75,0);
  \draw (2.25,-6) -- ++(.75,0);
  \draw (1,-7) -- ++(.75,0);
  \draw (2.25,-7) -- ++(.75,0);

  \draw (2,-6) circle [radius=.25] node {\scriptsize $f$};
  \draw (2,-7) circle [radius=.25] node {\scriptsize $g$};
  
  \draw (3,-7) -- ++(-.75,1);
  \draw (3,-6) -- ++(-.75,-1);
  
  \draw (1.75,-7.5) -- ++(.25,0) -- ++(0,.25);


  \draw (3,0) -- ++(.75,0);
  \draw (4.25,0) -- ++(.75,0);
  \draw[red,very thick] (3,-2) -- ++(.75,0) node [above,midway] {\scriptsize-$.85$};
  \draw[blue,very thick] (3,-2.1) -- ++(.75,0);
  \draw[red,very thick] (4.25,-2) -- ++(.75,0);
  \draw[blue,very thick] (4.25,-2.1) -- ++(.75,0);

  \draw (4,0) circle [radius=.25] node {\scriptsize $f$};
  \draw[red,very thick] (4,-2) circle [radius=.25] node {\scriptsize $g$};
  
  \draw (5,-2) -- ++(-.75,2);
  \draw[red,very thick] (5,0) -- ++(-.75,-2);
  \draw[blue,very thick] (5.1,-.1) -- ++(-.75,-2);
  
  \draw[red,very thick] (3.75,-2.5) -- ++(.25,0) -- ++(0,.25);
  \node[red] at (3.5,-2.5) {\scriptsize$0$};

  \draw (3,-1) -- ++(.75,0);
  \draw (4.25,-1) -- ++(.75,0);
  \draw[red,very thick] (3,-3) -- ++(.75,0) node [midway,fill=white] {\scriptsize$1.3$};
  \draw[red,very thick] (4.25,-3) -- ++(.75,0);

  \draw (4,-1) circle [radius=.25] node {\scriptsize $f$};
  \draw[red,very thick] (4,-3) circle [radius=.25] node {\scriptsize $g$};
  
  \draw (5,-3) -- ++(-.75,2);
  \draw[red,very thick] (5,-1) -- ++(-.75,-2);
  
  \draw[red,very thick] (3.75,-3.5) -- ++(.25,0) -- ++(0,.25);
  \node[red] at (3.5,-3.5) {\scriptsize$0$};

  \draw (3,-4) -- ++(.75,0);
  \draw (4.25,-4) -- ++(.75,0);
  \draw (3,-6) -- ++(.75,0);
  \draw (4.25,-6) -- ++(.75,0);

  \draw (4,-4) circle [radius=.25] node {\scriptsize $f$};
  \draw (4,-6) circle [radius=.25] node {\scriptsize $g$};
  
  \draw (5,-6) -- ++(-.75,2);
  \draw (5,-4) -- ++(-.75,-2);
  
  \draw (3.75,-6.5) -- ++(.25,0) -- ++(0,.25);

  \draw (3,-5) -- ++(.75,0);
  \draw (4.25,-5) -- ++(.75,0);
  \draw (3,-7) -- ++(.75,0);
  \draw (4.25,-7) -- ++(.75,0);

  \draw (4,-5) circle [radius=.25] node {\scriptsize $f$};
  \draw (4,-7) circle [radius=.25] node {\scriptsize $g$};
  
  \draw (5,-7) -- ++(-.75,2);
  \draw (5,-5) -- ++(-.75,-2);
  
  \draw (3.75,-7.5) -- ++(.25,0) -- ++(0,.25);


  \draw[red,very thick] (5,0) -- ++(.75,0) node [above,midway] {\scriptsize-$.77$};
  \draw[blue,very thick] (5.1,-.1) -- ++(.65,0);
  \draw[red,very thick] (6.25,0) -- ++(.75,0);
  \draw[blue,very thick] (6.25,-.1) -- ++(.75,0);
  \draw (5,-4) -- ++(.75,0);
  \draw (6.25,-4) -- ++(.75,0);

  \draw[red,very thick] (6,0) circle [radius=.25] node {\scriptsize $f$};
  \draw (6,-4) circle [radius=.25] node {\scriptsize $g$};
  
  \draw[red,very thick] (7,-4) -- ++(-.75,4);
  \draw (7,0) -- ++(-.75,-4);
  
  \draw (5.75,-4.5) -- ++(.25,0) -- ++(0,.25);

  \draw[red,very thick] (5,-1) -- ++(.75,0) node [above,midway] {\scriptsize$.56$};
  \draw[red,very thick] (6.25,-1) -- ++(.75,0);
  \draw (5,-5) -- ++(.75,0);
  \draw (6.25,-5) -- ++(.75,0);

  \draw[red,very thick] (6,-1) circle [radius=.25] node {\scriptsize $f$};
  \draw (6,-5) circle [radius=.25] node {\scriptsize $g$};
  
  \draw[red,very thick] (7,-5) -- ++(-.75,4);
  \draw (7,-1) -- ++(-.75,-4);
  
  \draw (5.75,-5.5) -- ++(.25,0) -- ++(0,.25);

  \draw[red,very thick] (5,-2) -- ++(.75,0) node [above,midway] {\scriptsize-$.08$};
  \draw[blue,very thick] (5,-2.1) -- ++(.75,0);
  \draw[red,very thick] (6.25,-2) -- ++(.75,0);
  \draw[blue,very thick] (6.25,-2.1) -- ++(.75,0);
  \draw (5,-6) -- ++(.75,0);
  \draw (6.25,-6) -- ++(.75,0);

  \draw[red,very thick] (6,-2) circle [radius=.25] node {\scriptsize $f$};
  \draw (6,-6) circle [radius=.25] node {\scriptsize $g$};
  
  \draw[red,very thick] (7,-6) -- ++(-.75,4);
  \draw (7,-2) -- ++(-.75,-4);
  
  \draw (5.75,-6.5) -- ++(.25,0) -- ++(0,.25);

  \draw[red,very thick] (5,-3) -- ++(.75,0) node [above,midway] {\scriptsize$.69$};
  \draw[red,very thick] (6.25,-3) -- ++(.75,0);
  \draw (5,-7) -- ++(.75,0);
  \draw (6.25,-7) -- ++(.75,0);

  \draw[red,very thick] (6,-3) circle [radius=.25] node {\scriptsize $f$};
  \draw (6,-7) circle [radius=.25] node {\scriptsize $g$};
  
  \draw[red,very thick] (7,-7) -- ++(-.75,4);
  \draw (7,-3) -- ++(-.75,-4);
  
  \draw (5.75,-7.5) -- ++(.25,0) -- ++(0,.25);
  
  \node at (7.75,.5) {\scriptsize\color{red}$\bm{y}$};
  \node at (7.75,0) {\scriptsize\color{red}$0.77$};
  \node at (7.75,-1) {\scriptsize\color{red}$0.56$};
  \node at (7.75,-2) {\scriptsize\color{red}-$0.08$};
  \node at (7.75,-3) {\scriptsize\color{red}-$1.51$};
  \node at (7.75,-4) {\scriptsize\color{red}-$1.44$};
  \node at (7.75,-5) {\scriptsize\color{red}$2.89$};
  \node at (7.75,-6) {\scriptsize\color{red}$2.33$};
  \node at (7.75,-7) {\scriptsize\color{red}-$0.69$};
  
  \node at (8.75,.5) {\scriptsize\color{blue}$\bm{e}_2$};
  \node at (8.75,0) {\scriptsize\color{blue}$1$};
  \node at (8.75,-1) {\scriptsize\color{blue}$0$};
  \node at (8.75,-2) {\scriptsize\color{blue}$1$};
  \node at (8.75,-3) {\scriptsize\color{blue}$0$};
  \node at (8.75,-4) {\scriptsize\color{blue}$0$};
  \node at (8.75,-5) {\scriptsize\color{blue}$0$};
  \node at (8.75,-6) {\scriptsize\color{blue}$0$};
  \node at (8.75,-7) {\scriptsize\color{blue}$0$};

\end{tikzpicture}
    \caption{Example of the proposed successive syndrome-check decoder for $\mathcal{P}(8,4)$ with $\mathcal{I} = \{u_3,u_5,u_6,u_7\}$ and $\mathcal{F} = \{u_0,u_1,u_2,u_4\}$. The LLR values received from the channel represent $\bm{x} = \{0,0,1,1,1,0,0,1\}$ with the syndrome $\bm{x}\bm{H} = \{0,0,1,0\}$. Therefore, the successive syndrome-check decoder starts by calculating the LLR value associated with $u_2$. The corresponding error vector is $\bm{e}_2 = \{1,0,1,0,0,0,0,0\}$ that is used to refine $\bm{x}$ to $\{1,0,0,1,1,0,0,1\}$. Since the syndrome of the refined $\bm{x}$ is $0$, the decoder outputs $\bm{u} = \bm{x}\bm{G}^{\otimes 3} = \{0,0,0,0,0,1,1,1\}$.}
    \label{fig:SSCdecEx}
\end{figure}

Fig.~\ref{fig:SSCdecEx} shows an example of the proposed successive syndrome-check decoder on the factor graph of $\mathcal{P}(8,4)$ with $\mathcal{I} = \{u_3,u_5,u_6,u_7\}$ and $\mathcal{F} = \{u_0,u_1,u_2,u_4\}$. The received values from the channel are turned into the LLR vector $\bm{y} = \{0.77,0.56,-0.08,-1.51,-1.44,2.89,2.33,-0.69\}$. The received codeword is then initialized based on (\ref{eq:xinit}) as $\bm{x} = \{0,0,1,1,1,0,0,1\}$. The syndrome of $\bm{x}$ is calculated as $\bm{x}\bm{H} = \{0,0,1,0\}$, which corresponds to $\{u_0,u_1,u_2,u_4\}$. Since $u_0 = u_1 = 0$, these two bits match the frozen bit value and thus, they do not need to undergo the decoding process. However, since $u_2 = 1$, the proposed successive syndrome-check decoder calculates its LLR value by traversing the factor graph from right to left, and stores the intermediate LLRs for error vector generation (see the red lines in Fig.~\ref{fig:SSCdecEx}). Note that the calculated LLR value of $u_2$ is indeed negative. The error vector $\bm{e}_2$ is then generated by traversing the factor graph from left to right (see the blue lines in Fig.~\ref{fig:SSCdecEx}). First, an $f$ function is encountered whose inputs are $-0.85$ and $1.3$. Therefore, the path corresponding to $-0.85$ is selected. Then, there is a $g$ function with inputs $-0.77$ and $-0.08$, and the path is extended on both inputs. Finally, the two paths are extended on the input of the two $f$ functions whose absolute LLR values are smaller. This results in $\bm{e}_2 = \{1,0,1,0,0,0,0,0\}$.

After $\bm{e}_2$ is generated, $\bm{x}$ is refined as
\begin{equation}
    \bm{x} = \bm{x} \oplus \bm{e}_2 = \{1,0,0,1,1,0,0,1\} \text{.}
\end{equation}
This refined codeword has the syndrome $\bm{x}\bm{H} = \{0,0,0,0\}$, which passes the syndrome check. Therefore, the proposed decoder terminates the decoding process and outputs $\bm{u} = \bm{x}\bm{G}^{\otimes 3} = \{0,0,0,0,0,1,1,1\}$.

Note that the proposed decoder only needs to generate an error vector for the bits whose LLR values do not match their bit estimation\footnote{An LLR value matches a bit estimation if the LLR value is negative and the bit estimation is $1$, or if the LLR value is positive and bit estimation is $0$.}. In SC decoding, each information bit is decoded by making a decision based on the sign of the LLR value. Therefore, in SC decoding, the LLR values of information bits always match their bit estimation. Consequently, the error vector generation is only needed for the violating frozen bits whose LLR values are negative.

\subsection{Extension to SCL Decoding}

In SCL decoding, each information bit is estimated as both $0$ and $1$. Therefore, if the calculated LLR value of an information bit for each candidate in the list does not match its bit estimation, an error vector needs to be generated. As a result, the proposed successive syndrome-check decoder can be adapted to SCL decoding by generating error vectors not only for the violating frozen bits, but also for information bits. Similar to the case of SC decoding, if the syndrome of $\bm{x}$ is not zero, the first $u_i\in\mathcal{F}$ with $u \neq 0$ is identified. However, if there are information bits before that frozen bit, error vectors are generated for each information bit, $\bm{x}$ is refined for each candidate in the list, and the syndrome is calculated again for all $\bm{x}$. The process is continued until the syndrome of the best path in the list becomes $0$. Note that since there is a list of candidates, the error vector is generated based on the first violating frozen bit throughout the list. This guarantees that the error-correction performance of the proposed decoder is equivalent to that of the SCL decoder.

\section{Results}

This section presents latency results for the proposed successive syndrome-check decoder in terms of the number of time steps and compares them with the latency of the SC and SCL decoders. Similar to the case of SC and SCL decoding, the latency of the successive syndrome-check decoder is calculated by considering a fully-parallel implementation. Moreover, each level traversal on the factor graph based on (\ref{eq:fFunction}) and (\ref{eq:gFunctionS}) is equivalent to $1$ time step, and the bit-wise operations in (\ref{eq:ferror}) and (\ref{eq:gerror}) to generate the error vector and the calculation of the syndrome do not incur additional latency.

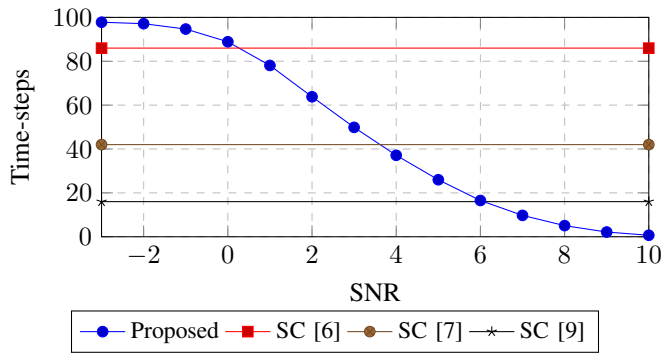
\begin{figure}[t]
    \centering
    \begin{tikzpicture}

\begin{axis}[
scale=1,
xmin=-3,
xmax=10,
ymin=0,
ymax=100,
ymajorgrids=true,
xmajorgrids=true,
grid style=dashed,
width=\linewidth, height=4.5cm,
xlabel={SNR},
ylabel={Time-steps},
legend cell align={left},
legend pos=north east,
legend style={
	column sep=0mm,
	font=\fontsize{9pt}{9}\selectfont,
},
legend to name=legend-PLatcomp,
legend columns=4,
]


\addplot
table {
-3 97.80
-2 97.13
-1 94.68
0 88.88
1 78.09
2 63.81
3 49.87
4 37.13
5 25.98
6 16.52
7 9.74
8 5.07
9 2.15
10 0.68
};
\addlegendentry{Proposed}

\addplot
table {
-3 86
10 86
};
\addlegendentry{SC \cite{alamdar2011simplified}}

\addplot
table {
-3 42
10 42
};
\addlegendentry{SC \cite{sarkis2014fast}}

\addplot
table {
-3 16
10 16
};
\addlegendentry{SC \cite{hanif2017fast}}

\end{axis}
\end{tikzpicture}
    \ref{legend-PLatcomp}
    \caption{Required number of decoding time-steps for $\mathcal{P}(128,64)$ used in the 5G standard. The FER at SNR~$=4$~dB is $1.6\times 10^{-3}$.}
    \label{fig:TvsSTime128}
\end{figure}

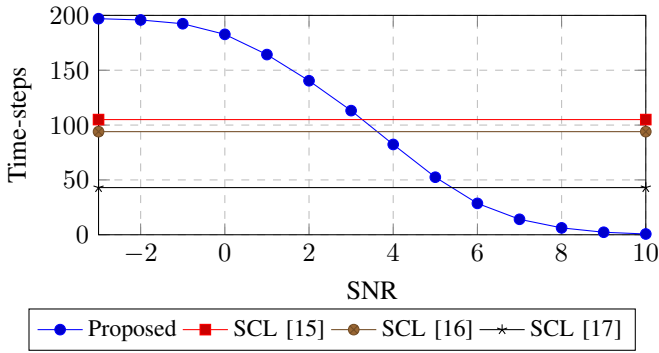
\begin{figure}[t]
    \centering
    \begin{tikzpicture}

\begin{axis}[
scale=1,
xmin=-3,
xmax=10,
ymin=0,
ymax=200,
ymajorgrids=true,
xmajorgrids=true,
grid style=dashed,
width=\linewidth, height=4.5cm,
xlabel={SNR},
ylabel={Time-steps},
legend cell align={left},
legend pos=north east,
legend style={
	column sep=0mm,
	font=\fontsize{9pt}{9}\selectfont,
},
legend to name=legend-PLatcompSCL,
legend columns=4,
]


\addplot
table {
-3 197.00
-2 195.83
-1 192.39
0  182.71
1  164.27
2  140.40
3  113.17
4  82.320
5  52.460
6  28.590
7  14.090
8  6.2700
9  2.2800
10 0.6900
};
\addlegendentry{Proposed}

\addplot
table {
-3 105
10 105
};
\addlegendentry{SCL \cite{hashemi2016fast}}

\addplot
table {
-3 94
10 94
};
\addlegendentry{SCL \cite{hashemi2017fast}}

\addplot
table {
-3 43
10 43
};
\addlegendentry{SCL \cite{hanif2018fast}}

\end{axis}
\end{tikzpicture}
    \ref{legend-PLatcompSCL}
    \caption{Required number of decoding time-steps for $\mathcal{P}(128,64)$ used in the 5G standard. Note that $L=8$ for the SCL decoders. The FER at SNR~$=4$~dB is $1.1\times 10^{-3}$.}
    \label{fig:TvsSTime128SCL}
\end{figure}

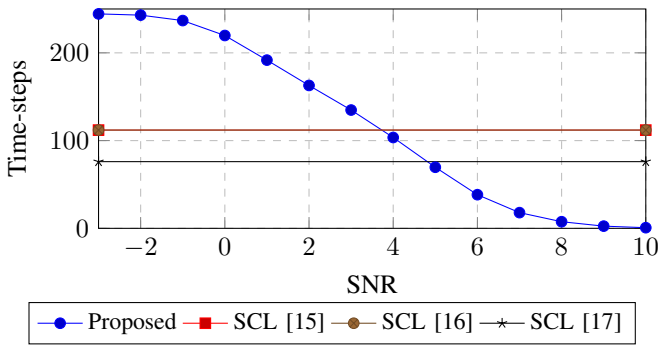
\begin{figure}[t]
    \centering
    \begin{tikzpicture}

\begin{axis}[
scale=1,
xmin=-3,
xmax=10,
ymin=0,
ymax=250,
ymajorgrids=true,
xmajorgrids=true,
grid style=dashed,
width=\linewidth, height=4.5cm,
xlabel={SNR},
ylabel={Time-steps},
legend cell align={left},
legend pos=north east,
legend style={
	column sep=0mm,
	font=\fontsize{9pt}{9}\selectfont,
},
legend to name=legend-RMLatcomp,
legend columns=4,
]


\addplot
table {
-3 244.39
-2 242.99
-1 236.79
0  219.77
1  191.72
2  162.85
3  134.82
4  103.53
5  69.610
6  38.330
7  17.890
8  7.5500
9  2.5100
10 0.7300
};
\addlegendentry{Proposed}

\addplot
table {
-3 112
10 112
};
\addlegendentry{SCL \cite{hashemi2016fast}}

\addplot
table {
-3 112
10 112
};
\addlegendentry{SCL \cite{hashemi2017fast}}

\addplot
table {
-3 76
10 76
};
\addlegendentry{SCL \cite{hanif2018fast}}

\end{axis}
\end{tikzpicture}
    \ref{legend-RMLatcomp}
    \caption{Required number of decoding time-steps for a RM code of length $128$ and rate $\frac{1}{2}$. Note that $L=32$ for the SCL decoders. The FER at SNR~$=4$~dB is $2.3\times 10^{-5}$.}
    \label{fig:TvsSTime128RM}
\end{figure}

Fig.~\ref{fig:TvsSTime128} presents the required number of decoding time-steps for the proposed successive syndrome-check decoder that is based on SC decoding for $\mathcal{P}(128,64)$ used in the 5G standard. The results are compared with three fast SC decoders in the literature \cite{alamdar2011simplified,sarkis2014fast,hanif2017fast}. Fig.~\ref{fig:TvsSTime128SCL} shows the latency results of the proposed successive syndrome-check decoder based on SCL decoding for $\mathcal{P}(128,64)$ used in the 5G standard. The results are compared with three fast SCL decoders in \cite{hashemi2016fast,hashemi2017fast,hanif2018fast} and for all the decoders, $L=8$. Fig.~\ref{fig:TvsSTime128RM} plots the number of time-steps required for the proposed decoders and the fast SCL decoders in \cite{hashemi2016fast,hashemi2017fast,hanif2018fast} to decode a Reed-Muller (RM) code of length $128$ and rate $\frac{1}{2}$. The list size is set to $L=32$ to provide a frame error rate (FER) close to the maximum-likelihood decoding performance. It can be seen that in all the considered cases, the proposed method requires a smaller number of decoding time-steps in comparison with all the considered fast SC and SCL decoders at high signal-to-noise ratio (SNR) values. Note that in each considered scenario, the FER performance of the decoders are identical.

\section{Summary and Future Work}

In this paper, a successive syndrome-check decoder is presented that allows for early termination in SC and SCL decoding of polar codes. The proposed decoder is lossless in the sense that it does not degrade the error-correction performance of the underlying SC or SCL decoder. Moreover, the generation of error vectors in the proposed decoder obviates the need for generating partial sums, which are required in SC and SCL decoding. Simulation results show that at the same FER, the latency of the proposed decoder is smaller than that of the SC and SCL decoders at high SNR values.

Future work includes adapting the proposed successive syndrome-check decoder to the fast SC and SCL decoders. This ensures the worst-case latency of the decoder is the latency of the corresponding fast SC or SCL decoder. Another interesting direction for future work is to use a cyclic redundancy-check (CRC) to enable early-termination in addition to checking the syndrome. This is particularly useful in applications where polar codes are concatenated with a CRC.

\section*{Acknowledgment}

This work is supported in part by ONR grant N00014-18-1-2191. S.~A.~Hashemi was supported by a Postdoctoral Fellowship from the Natural Sciences and Engineering Research Council of Canada (NSERC) and by Huawei. M. Mondelli was partially supported by the 2019 Lopez-Loreta Prize.


\end{document}